\documentclass{emulateapj}

\usepackage[latin1]{inputenc}
\usepackage{amsmath}
\usepackage{amsfonts}
\usepackage{amssymb}
\usepackage{bm}
\usepackage{default}
\usepackage{float}
\usepackage{hyperref}
\usepackage{graphicx}
\usepackage{gensymb}
\usepackage{subfigure}
\usepackage{units}

\newcommand{\cross}{\bm{\times}}
\newcommand{\dotprod}{\bm{\cdot}}

\renewcommand{\vec}[1]{\bm{#1}}

\shorttitle{Directed Energy NEO Deflection Simulations}
\shortauthors{Zhang et al.}
\submitted{}
\journalinfo{Accepted by PASP}

\begin{document}

\title{Orbital Simulations on Deflecting Near-Earth Objects by Directed Energy}

\author{Qicheng Zhang\altaffilmark{1}, Kevin J. Walsh\altaffilmark{2}, Carl Melis\altaffilmark{3}, Gary B. Hughes\altaffilmark{4}, Philip M. Lubin\altaffilmark{1}}
\email{qicheng@cometary.org}

\affil{$^1$ Dept. of Physics, Univ. of California, Santa Barbara, CA USA 93106-9530\\
$^2$ Southwest Research Institute, Boulder, CO USA 80302\\
$^3$ Center for Astrophysics and Space Sciences, Univ. of California, San Diego, CA USA 92093-0424\\
$^4$ Statistics Dept., California Polytechnic State Univ., San Luis Obispo, CA USA 93407-0405}

\begin{abstract}
Laser ablation of a Near-Earth Object (NEO) on a collision course with Earth produces a cloud of ejecta which exerts a thrust on the NEO, deflecting it from its original trajectory. Ablation may be performed from afar by illuminating an Earth-targeting asteroid or comet with a \emph{stand-off} ``DE-STAR'' system consisting of a large phased-array laser in Earth orbit. Alternatively, a much smaller \emph{stand-on} ``DE-STARLITE'' system may travel alongside the target, slowly deflecting it from nearby over a long period. This paper presents orbital simulations comparing the effectiveness of both systems across a range of laser and NEO parameters. Simulated parameters include magnitude, duration and, for the stand-on system, direction of the thrust, as well as the type, size and orbital characteristics of the target NEO. These simulations indicate that deflection distance is approximately proportional to the magnitude of thrust and to the square of the duration of ablation, and is inversely proportional to the mass. Furthermore, deflection distance shows strong dependence on thrust direction with the optimal direction of thrust varying with the duration of laser activity. As one example, consider a typical \unit[325]{m} asteroid: beginning \unit[15]{yr} in advance, just \unit[2]{N} of thrust from a $\unit[\sim20]{kW}$ stand-on DE-STARLITE system is sufficient to deflect the asteroid by $\unit[2]{R_{\earth}}$. Numerous scenarios are discussed as is a practical implementation of such a system consistent with current launch vehicle capabilities.
\end{abstract}

\section{Introduction}

A wide array of concepts for the deflection of threatening Near-Earth Objects (NEO) have been proposed. Several detailed surveys of threat mitigation strategies are available such as \citet{belton:2004}, \citet{gritzner:2004}, \citet{colombo:2009}, \citet{cuartielles:2007} and \citet{morrison:2002}. These strategies fall into several categories, including, but not limited to

\begin{enumerate}
\item Kinetic impactors, with or without explosive charges. An expendable spacecraft would be sent to intercept the threatening object. Direct impact would modify the object's orbit through momentum transfer. Enhanced momentum transfer can occur using an explosive charge such as a nuclear weapon  \citep{koenig:2007,melosh:1997,mcinnes:2004,conway:2004}.
\item Gradual orbit deflection by surface albedo alteration. The albedo of an object could be changed using paint \citep{hyland:2010}, mirrors \citep{vasile:2010}, sails \citep{maddock:2007}, etc. As the albedo is altered, a change in the object's Yarkovsky thermal drag would gradually shift the object's orbit.
\item Direct motive force, such as by mounting a thruster directly to the object. Thrusters could include chemical propellants, solar or nuclear powered electric drives, or ion engines \citep{walker:2005}. A reversed setup is also possible where a ``shepherd'' spacecraft directs of a beam of high-speed ions to collide with, and thus, transfer momentum to the asteroid \citep{bombardelli:2011}.
\item Indirect orbit alteration, such as by gravity tractors. A spacecraft with sufficient mass would be positioned near the object, and maintain a fixed station with respect to the object using on-board propulsion. Gravitational attraction would tug the object toward the spacecraft, and gradually modify the object's orbit \citep{schweikart:2006}.
\item Expulsion of surface material such as by robotic mining. A robot on the surface of an object would repeatedly eject material from the object. The reaction force by the ejected material alters the object's trajectory \citep{olds:2007}.
\item Vaporization of surface material. Like robotic mining, vaporization on the surface of an object continually ejects the vaporized material, creating a reactionary force that pushes the object into a new path. Vaporization can be accomplished by solar concentrators \citep{gibbings:2011} or lasers \citep{maddock:2007} deployed on spacecraft stationed near the asteroid. One study envisioned a single large reflector mounted on a spacecraft traveling alongside an asteroid \citep{kahle:2006}. The idea was expanded to a formation of spacecraft orbiting in the vicinity of the asteroid, each equipped with a smaller concentrator assembly capable of focusing solar power onto an asteroid at distances near $\unit[\sim1]{km}$ \citep{vasile:2010}. Efficiency of a laser system for surface ablation can be enhanced using an array of phase-locked lasers \citep{kosmo:2014:spie-op}, allowing more photonic flux to be delivered to the asteroid and at greater distances. Envisioning ever larger arrays of phase-locked lasers allows contemplation of stand-off systems that could deliver sufficient flux to the surface of a distant NEO from Earth orbit \citep{lubin:2014:oe}.
\end{enumerate}

Simulations were developed in order to measure the effectiveness of deflection of a threat by laser ablation as proposed in \citet{kosmo:2014:spie-op} and \citet{lubin:2014:oe}. Both stand-off and stand-on missions are discussed. The much larger stand-off system (called \emph{DE-STAR} for \emph{Directed Energy System for Targeting of Asteroids and exploRation}) consists of a laser which remains in Earth orbit ablating the target from afar. The smaller (``lite'') stand-on system (called DE-STARLITE) involves a laser being physically delivered to the target. The laser technology is described in much greater detail in \citet{lubin:2013:spie-op}, \citet{kosmo:2014:spie-op} and \citet{lubin:2014:oe}. Effects of asteroid rotation are discussed in \citet{johansson:2014:spie-op} and optical modeling is discussed in \citet{hughes:2013:spie-op} and \citet{hughes:2014:spie-op}.

Emphasis is placed on the more practical stand-on system which can be built more rapidly and inexpensively as a near-term solution due to its reduced scale. However, the full stand-off system is still considered as a possibility for the more distant future for its ability to rapidly respond to identified threats and, moreover, to target objects like long-period comets in orbits unreachable by current propulsion technology.

\subsection{Laser Ablation of an Object's Surface}

The objective of the laser directed energy system is to project a large enough flux onto the surface of the asteroid to heat the surface to a temperature that exceeds the vaporization point of constituent materials, typically $\unit[\sim2\,500]{K}$, corresponding to a flux $\unit[\sim10]{MW\cdot m^{-2}}$. The reactionary thrust of the ejecta plume will divert the asteroid's trajectory.

To produce sufficient flux, the system must have both adequate beam convergence and sufficient power. Optical aperture size, pointing control and jitter, and efficacy of adaptive optics techniques are several critical factors that affect beam convergence. The optical power output of a stand-on DE-STARLITE mission can be varied depending on the target size and warning time and might range from $\unit[\sim1]{kW}$ to $\unit[\sim1]{MW}$. Current laser electrical-to-optical ``wallplug efficiency'' baselines are nearing 50\%. Even higher efficiency allows for more thrust on the target for a given electrical input as well as for smaller radiators and hence lower mission mass \citep{kosmo:2014:spie-op}. For the large-scale stand-off systems considered in this paper, a total solar-to-laser optical power efficiency of 50\% is assumed.

Evaporation at the laser spot produces a vaporization plume thrust that can be used to change the asteroid's orbit and effectively deflect asteroids from colliding with Earth. In \citet{lubin:2014:oe}, \citet{johansson:2014:spie-op} and \citet{lubin:2013:spie-op}, simulations are performed with the high temperature materials expected in rocky target that require the highest flux and the low temperature volatiles in comets that can also be deflected with much less flux. This paper assumes a conversion factor of $\unit[100]{\mu N\cdot W^{-1}}$, as expected from thermal simulations and measurements, in orbital simulations of various NEO deflection scenarios.

\vspace{7mm}
\section{Orbital Simulations}

The simulation considers the 3-body system consisting of the Sun, Earth and NEO. The Moon is not considered as a separate body, but its mass is combined with that of the Earth. This combined ``Earth-Moon point mass'' is denoted here simply as the Earth. The objects are numerically integrated as an 3-body system of mutually gravitating point masses.

The acceleration of the NEO is divided into two components:

\begin{equation}
\vec{a}=\vec{a_\text{g}}+\vec{a_\text{l}}
\end{equation}

The first component $\vec{a_\text{g}}$ is the net gravitational acceleration from the Sun and Earth which are integrated together as part of the same 3-body simulation. The second component $\vec{a_\text{l}}\equiv a_\text{l}\vec{\hat{a_\text{l}}}$ is a  perturbation of the NEO by the laser's thrust $F=ma_\text{l}$ for a NEO of mass $m$. For these simulations, the NEO is assumed to be spherical with a uniform density of $\rho=\unit[2\,000]{kg\cdot m^{-3}}$ for an asteroid and $\rho=\unit[600]{kg\cdot m^{-3}}$ for a comet. The direction of thrust $\vec{\hat{a_\text{l}}}$ varies depending on the mode by which the thrust is applied.

\subsection{Stand-On Mode}

In the \emph{stand-on} thrust case, the laser is maneuvered in close proximity to the target NEO. Due to the difficulty of delivering a massive spacecraft into an orbit typical of most comets, only asteroids are considered as targets for stand-on missions.

The stand-on laser system may be placed a distance $\delta\sim\unit[10]{km}$ either ahead or behind the asteroid in its orbit depending on the desired direction of thrust. This distance is sufficiently large for the asteroid's gravity to have limited effect on the laser's trajectory. Even a large \unit[400]{m} asteroid produces a perturbation $\unit[\sim4\times10^{-8}]{m\cdot s^{-2}}$, a minuscule acceleration similar in magnitude (and opposite in direction) to the photon-imparted acceleration by a \unit[10]{kW} laser beam on a \unit[1\,000]{kg} spacecraft. Note that $\delta\propto v$ where $v$ is the heliocentric speed of the asteroid and laser system, and $v\propto\sqrt{1/r}$ where $r$ is their distance from the Sun. For the low eccentricity ($e\lesssim0.5$) orbits considered, variations in $\delta$ will be small and are at most restricted to $\pm25\%$ of the nominal distance for $e=0.5$.

\begin{figure}[H]
\plotone{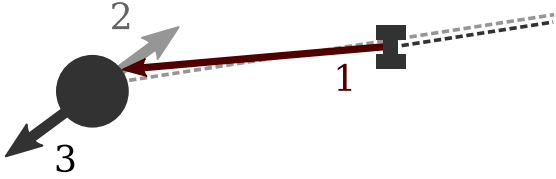}
\caption{A stand-on system (right) trails the target asteroid (left) in solar orbit (dotted lines). The laser beam (1) heats and vaporizes material at a spot on the asteroid, producing an ejecta plume (2) which acts as a propellant, exerting a thrust (3) on the asteroid. An $\alpha=\beta=0\degree$ thrust is obtained when the laser beam (1) is centered on the asteroid in this configuration, producing a plume (2) opposite the asteroid's velocity, yielding a thrust (3) on the asteroid parallel to its velocity. A thrust with $\alpha>90\degree$ requires the laser be positioned ahead of the asteroid in its orbit.}
\label{fig:diagram}
\end{figure}

At $\delta=\unit[10]{km}$, a \unit[1]{m} phased array produces a $\unit[\sim1]{cm}$ laser spot on the asteroid which is sufficiently small for \unit[1]{kW} to ablate material at a flux of $\unit[\sim10]{MW\cdot m^{-2}}$. Alternatively, with a more powerful system of at least \unit[100]{kW}, a similar flux at the laser spot may be achieved by simply focusing the beam(s) with a \unit[10]{cm} lens. The location of the laser spot on the asteroid, marking the site at which ablation occurs, may be selected to be anywhere on the spacecraft-facing side of the asteroid, so the generated thrust may be selected to be in nearly any direction. Figure \ref{fig:diagram} illustrates the relation between the direction of the laser beam, the produced ejecta plume and the resulting thrust that is exerted on the asteroid.

This model considers the special case where the direction of thrust $\vec{\hat{a_\text{l}}}$ is fixed relative to the direction of the asteroid's velocity $\vec{\hat{v}}$ and that of its orbital momentum $\vec{\hat{l}}\equiv\vec{\hat{r}}\cross\vec{\hat{v}}$. Thrust direction $\vec{\hat{a_\text{l}}}$ may then be specified in the frame defined by these directions by an azimuth angle $\alpha$ and elevation angle $\beta$ as given by Equation \ref{eq:lasaccel_standon}.

\begin{equation}
\vec{\hat{a_\text{l}}}=\vec{\hat{v}}\cos\alpha\cos\beta+\left(\vec{\hat{l}}\cross\vec{\hat{v}}\right)\sin\alpha\cos\beta+\vec{\hat{l}}\sin\beta
\label{eq:lasaccel_standon}
\end{equation}

The magnitude of thrust on the asteroid is then simply $F=F_0=ma_\text{l}$.

\subsection{Stand-Off Mode}
\label{sec:standoff}

In the \emph{stand-off} thrust case, the laser is a satellite in orbit around the Earth, both of which are considered to be at a common heliocentric position $\vec{r}_{\earth}$. From a distance, the laser ablates material off the Earth-facing side of the target NEO at heliocentric position $\vec{r}$. This material, ejected toward the Earth, exerts a thrust on the NEO in the opposite direction, away from Earth. Therefore, the thrust on the NEO must be in the direction of its geocentric position vector:

\begin{equation}
\vec{\hat{a_\text{l}}}=\frac{\vec{r}-\vec{r}_{\earth}}{\|\vec{r}-\vec{r}_{\earth}\|}
\label{eq:lasaccel_standoff}
\end{equation}

Because the laser operates at a large distance from the target, the laser beam must diverge due to diffraction effects. At a distance $\Delta\equiv\|\vec{r}-\vec{r}_{\earth}\|$, a phased-array laser of diameter $d$ produces a spot roughly of diameter

\begin{equation}
D_\text{spot}=\frac{2\lambda}{d}\Delta
\end{equation}

Approximating the spot illumination as uniform and thrust as proportional to incident power, the thrust on a target of diameter $D$ is

\begin{equation}
F=F_0\times
\begin{cases}
1 & \text{ if } D_\text{spot}\leq D\\
\left(D/D_\text{spot}\right)^2 & \text{ if } D_\text{spot}>D
\end{cases}
\end{equation}

where $F_0\propto P$ is the thrust produced by ablation with the full power $P$ of the laser.

Note, however, that for laser ablation and thus significant thrust generation to occur, $D_\text{spot}$ must be smaller than some power and target-dependent $D_\text{crit}$. When $D_\text{spot}>D_\text{crit}$, there is insufficient flux density to raise the temperature on the target to its vaporization temperature $T_\text{crit}$ and thus activate the ablation process. $T_\text{crit}$ is only reached when $\Delta$ is below some critical distance $\Delta_\text{crit}$. To estimate $\Delta_\text{crit}$, the target is approximated as a perfect blackbody with radiation being the only mode of transport for thermal energy. Then,

\begin{equation}
\Delta_\text{crit}=\frac{d}{\lambda}\sqrt{\frac{P}{\pi\sigma T_\text{crit}^4}}
\end{equation}

In addition, to prevent cancellation of thrust over time, the laser should only be activated for a consistent sign of the quantity

\begin{equation}
\xi=\left(\vec{r}-\vec{r}_{\earth}\right)\dotprod\vec{v}
\end{equation}

where $\vec{v}$ is the heliocentric velocity of the target NEO. The sign of $\xi$ defines whether the Earth is \emph{ahead} ($\xi<0$) or \emph{behind} ($\xi>0$) the NEO in its orbit which determines whether thrust from the laser advances or delays the motion of the NEO respectively.

The mean power $\bar{P}$ output by the laser array over an extended period is constrained by the power output of its solar array. Unless otherwise stated, a stand-off laser array of diameter $d$ is assumed to be accompanied by a square solar array of side length $d$ operating at 50\% efficiency in Earth orbit, giving

\begin{equation}
\label{eq:arraypow}
\bar{P}=0.5 J d^2
\end{equation}

where $J=\unit[1360]{W\cdot m^{-2}}$ is used as the solar flux incident on the solar array. 

Note that Equation \ref{eq:arraypow} assumes constant direct solar illumination which is not necessarily the case, particularly for satellites in low-Earth orbit where Earth's shadow might shade a substantial fraction of the orbit. This shading problem may be minimized by placing the laser in a higher altitude orbit in a dawn/dusk sun-synchronous configuration.

Furthermore, the laser is not necessarily able to target the NEO continuously. The light path between the laser and the NEO may be interrupted by the Earth for a fraction of the laser's orbit around the Earth, preventing the laser beam from reaching the NEO. However, the $\bar{P}$ constrained by the energy budget provided by the solar array is the \emph{mean} power of the laser. The instantaneous power $P$ of the laser at any given time is \emph{not} constrained by this energy budget and is, instead, constrained by the laser elements in the array, which for these simulations, is assumed to be capable of producing a maximum total power $P_{\text{max}}\gg \bar{P}$ given a sufficient reservoir of energy. Therefore, with the support of a sufficiently large and efficient battery system, the mean power delivered to the NEO can be maintained at nearly $\bar{P}$ as given by Equation \ref{eq:arraypow} by switching between an instantaneous laser power of $P=0$ (charging the battery system with power $\bar{P}$) when the NEO is obstructed by Earth and $P=P_0>\bar{P}$ (drawing the excess power $P_0-\bar{P}$ from the battery) when targeting of the NEO is possible. As cycling between $P=0$ and $P=P_0$ occurs rapidly relative to the NEO's motion through the solar system, the assumed relation $F_0\propto P$ is well-approximated by $\bar{F_0}\propto\bar{P}$ with the same constant of proportionality. Therefore, rather than flicker between $P=0$ and $P_0$, the simulations simply consider $\bar{P}\to P$ and $\bar{F_0}\to F_0$ which yields nearly the same long-term dynamics given the existing assumptions.

Also, because $\Delta_\text{crit}\propto\sqrt{P}$, a stand-off array with a battery system could, in theory, extend its $\Delta_\text{crit}$ considerably by activating the array at $P=P_{\max}$ for short periods and directing the solar array to charge the battery in the remaining time. A detailed analysis on the feasibility and practical concerns of constructing and using a battery-supported laser array is beyond the scope of this paper and is a topic for future discussion.

\subsection{Initial Conditions Generation}

The orbit of a target NEO may be characterized by three parameters: semi-major axis ($a$), eccentricity ($e$) and inclination to the ecliptic ($i$). Its intersection with Earth in space and time constrains the remaining three degrees of freedom. These simulations consider an intersection at Earth's aphelion. However, due to the near circular shape of Earth's orbit, simulation results are nearly identical for other points of intersection. In the full 3-body system considered, these parameters are not constants of motion. The $a$, $e$ and $i$ identified for each simulation are the heliocentric orbital elements of the NEO prior to impact, before the NEO enters the Earth's gravitational influence, but may be very different earlier at the time $t=0$ for which initial conditions are computed.

Initial conditions of a NEO with a trajectory fitting a set of desired parameters are generated by the following procedure:

\begin{enumerate}
\item Let time of impact be designated $t=T$ and be defined as the time when the NEO and Earth occupy the same heliocentric position $\vec{r}(T)$, the position of impact.
\item Neglecting the gravity of Earth for this step only, use the desired $a$, $e$, $i$ to fit a 2-body heliocentric trajectory $\vec{\tilde{r}}(t)$ for the NEO through $\vec{r}(T)$. Compute $\vec{\tilde{r}}(T-\delta t)$ and $\vec{\tilde{v}}(T-\delta t)$, the position and velocity of the NEO a time $\delta t\sim\unit[1]{d}$ prior to impact (in the 2-body system).
\item Neglecting the NEO for this step only, fit a 2-body heliocentric trajectory $\vec{\tilde{r}}_{\earth}(t)$ for the Earth through $\vec{r}(T)$. Compute $\vec{\tilde{r}}_{\earth}(T-\delta t)$ and $\vec{\tilde{v}}_{\earth}(T-\delta t)$ of the Earth.
\item In a full 3-body system, use $\vec{\tilde{r}}(T-\delta t),\vec{\tilde{v}}(T-\delta t)$ for the NEO and $\vec{\tilde{r}}_{\earth}(T-\delta t)$, $\vec{\tilde{v}}_{\earth}(T-\delta t)$ for the Earth which avoids the singularity at $t=T$ where the two gravitational sources coincide. Finally, numerically integrate the time-reversed system to $t=0$ to obtain the initial conditions for the NEO ($\vec{r}(0)$, $\vec{v}(0)$) and those of the Earth ($\vec{r}_{\earth}(0)$, $\vec{v}_{\earth}(0)$).
\end{enumerate}

The NEO ($\vec{r}(0)$, $\vec{v}(0)$) and the Earth ($\vec{r}_{\earth}(0)$, $\vec{v}_{\earth}(0)$) are then integrated forward together with the Sun through the NEO's encounter with Earth under the perturbed 3-body system described earlier.

\section{Deflection Simulation Results}

Deflection of threatening NEO using both stand-off thrust, provided by a DE-STAR system, and stand-on thrust, provided by a DE-STARLITE system, was considered for a range of NEO sizes and orbits. Simulations were performed with a standard solar system N-body integrator package, SyMBA, using the mixed variable symplectic mapping (MVS) integrator \citep{duncan:1998}.

The effectiveness of a given setup with a particular target NEO is measured by the \emph{miss distance} of the NEO to the Earth. Miss distance (or alternatively, \emph{deflection distance}) is defined to be $\Delta_{\text{min}}$ which is computed as the nearest local minimum of the function $\Delta(t)\equiv\|\vec{r}(t)-\vec{r}_{\earth}(t)\|$ to $t=T$, the time of impact for the unperturbed NEO.

The asteroid 99942 Apophis is a well-known case of a Potentially Hazardous Object. It is a relatively large Atens group asteroid with a diameter of approximately \unit[325]{m} with an orbit of semi-major axis $a=\unit[0.92]{au}$, eccentricity $e=0.19$ and inclination $i=3.3\degree$. These orbital parameters are used here for the canonical orbit of a near-Earth asteroid.

\subsection{Stand-On Results}

\begin{figure*}
\plottwo{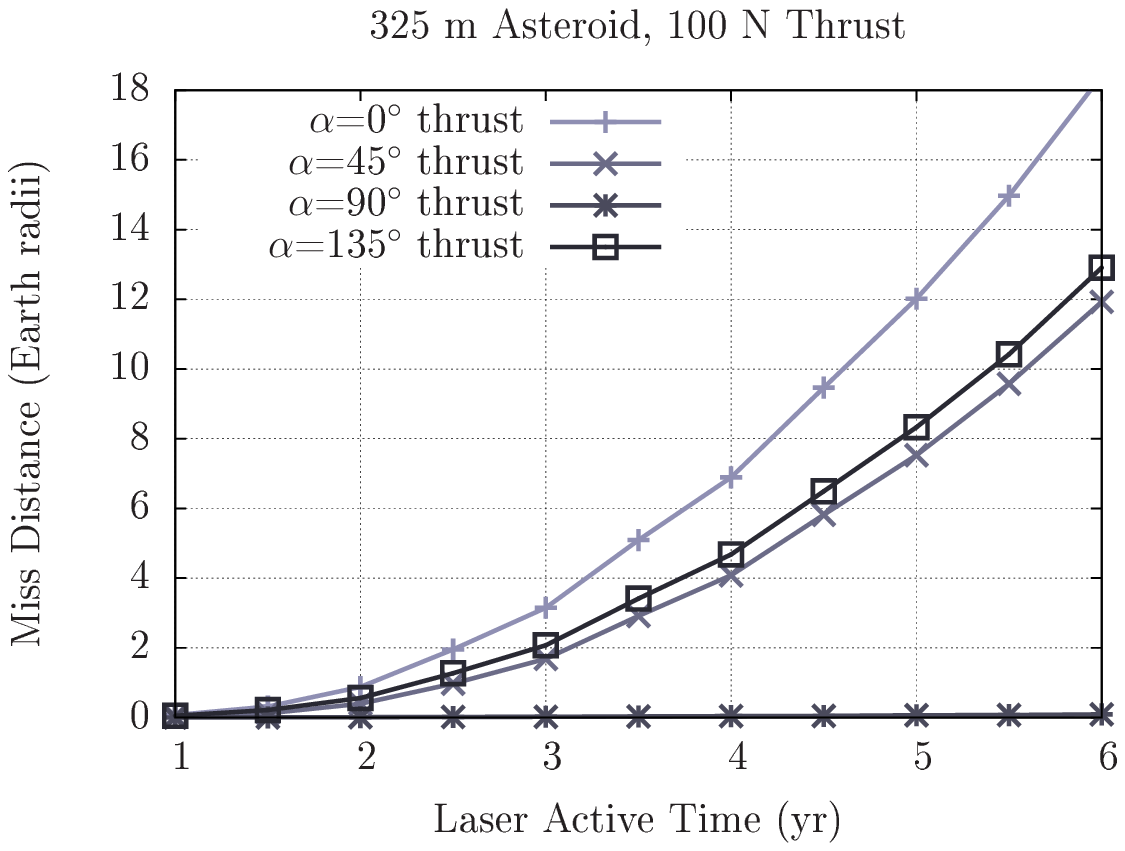}{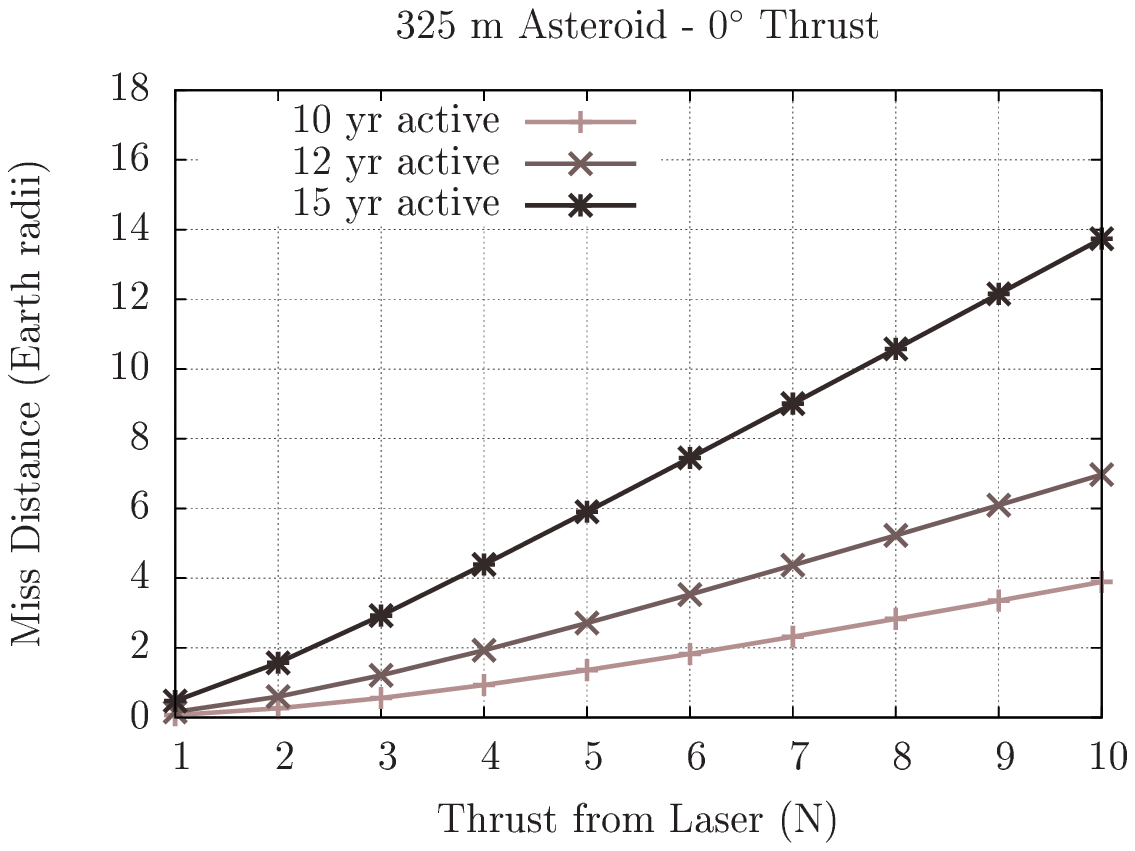}
\caption{For \unit[100]{N} applied to a \unit[325]{m} asteroid (left), deflection distance increases roughly quadratically with increasing time of laser activity. Thrust directed parallel to the asteroid's velocity vector is more effective than thrust directed $45\degree$, $90\degree$ and $135\degree$ from the velocity vector in the plane of the asteroid's orbit. Significantly less thrust of $< \unit[10]{N}$ in the parallel to velocity direction is needed to deflect the asteroid by $\unit[2]{R_{\earth}}$ if the available time for laser activity is increased to 10-\unit[15]{yr} (right). Deflection distance varies approximately linearly with thrust.}
\label{fig:325m-ast}
\end{figure*}

For a stand-on mission, it is conceivable to achieve a thrust of up to $F=\unit[100]{N}$ with a $\unit[\sim1]{MW}$ laser. Such a thrust may deflect the \unit[325]{m} asteroid to a miss distance of $\unit[2]{R_{\earth}}$ in as little as \unit[2.5]{yr} with thrust in the $\alpha=\beta=0\degree$ direction (``0\degree thrust'') -- the direction of the asteroid's velocity -- which appears to be near the optimal direction for a laser active over several years. With a decade of laser activity, deflection to $\unit[2]{R_{\earth}}$ is possible with less than \unit[7]{N} thrust. Given \unit[15]{yr} of ablation, \unit[2.5]{N} thrust is sufficient. In each case, besides a gravitational deviation by Earth at small thrust, miss distance grows roughly quadratically with increased time and linearly with increased thrust as seen in Figure \ref{fig:325m-ast}.

\begin{figure*}
\plottwo{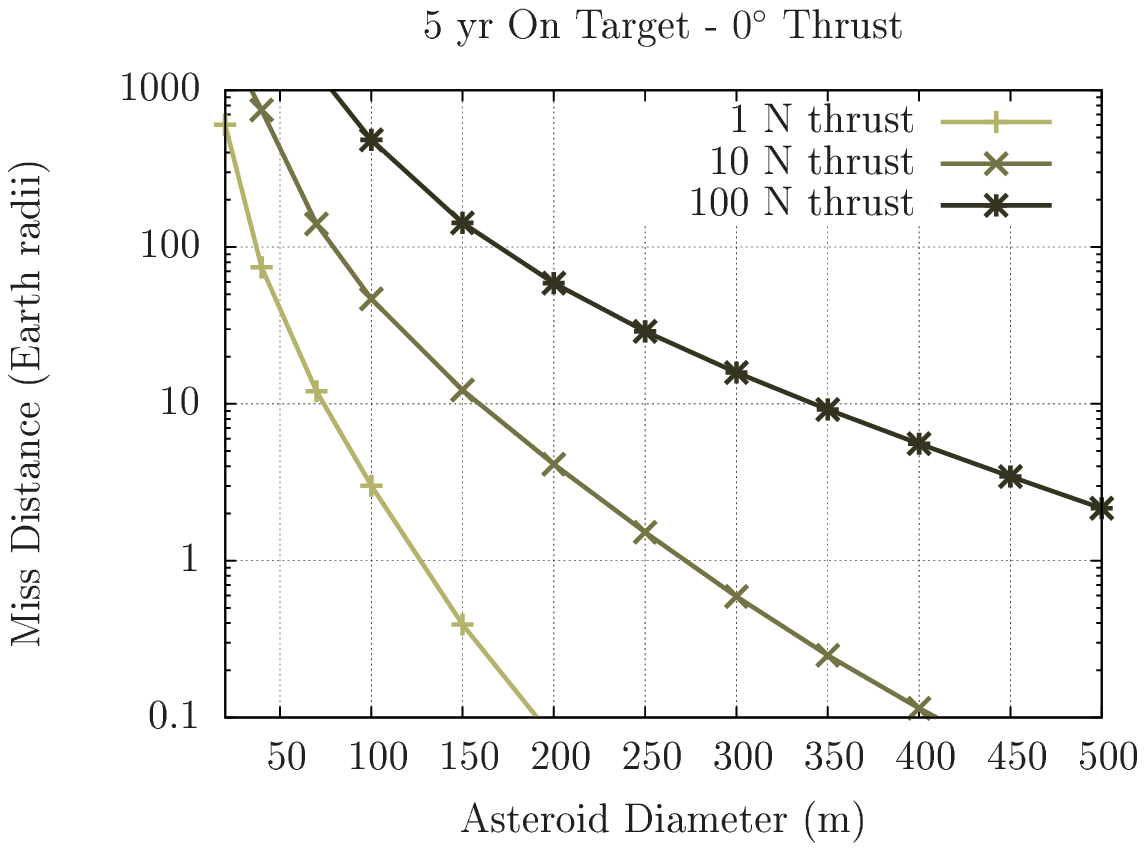}{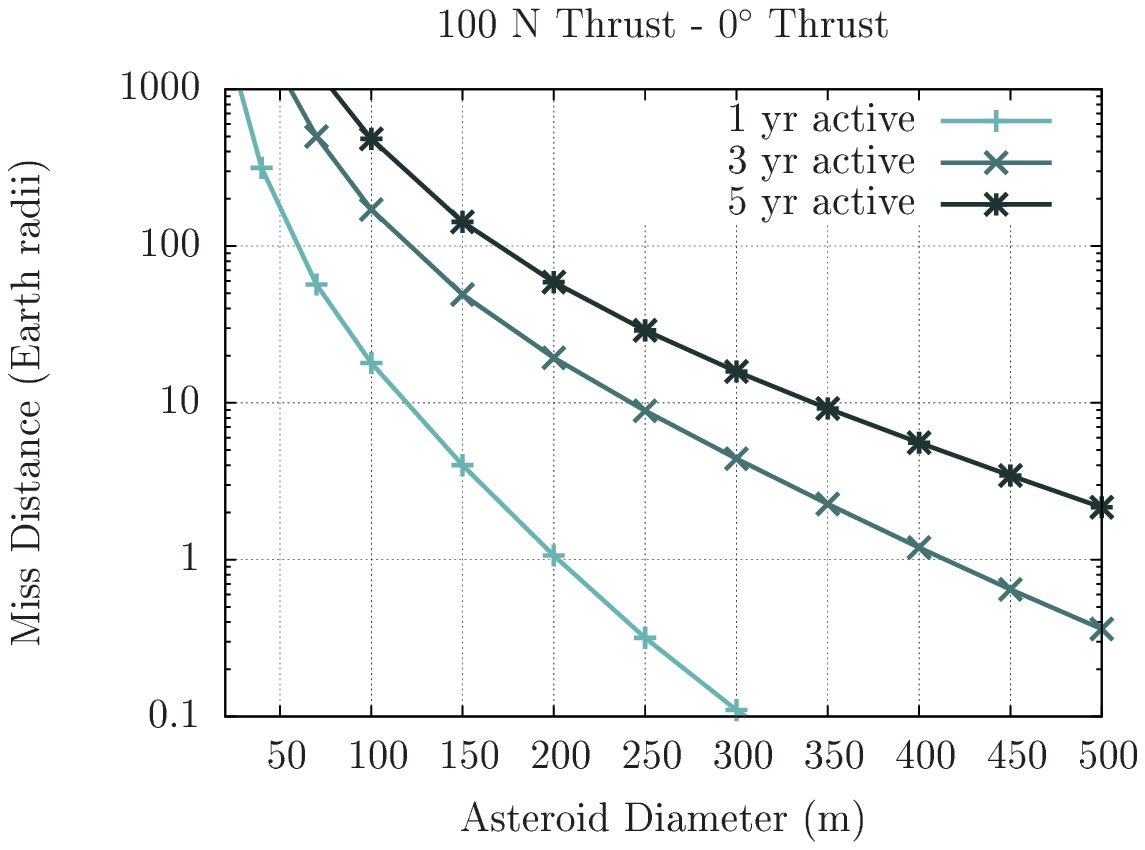}
\caption{Miss distance scales inversely with the mass of the target asteroid and thus the cube of its diameter, with the assumed uniform density. With \unit[5]{yr} of deflection (left), a \unit[500]{m} asteroid may be deflected by $\unit[2]{R_{\earth}}$ with \unit[100]{N} thrust, a \unit[230]{m} asteroid with \unit[10]{N} thrust or a \unit[100]{m} asteroid with \unit[1]{N} thrust. A large \unit[100]{N} stand-on system (right) can also deflect a \unit[350]{m} asteroid in \unit[3]{yr} or a \unit[170]{m} asteroid in \unit[1]{yr} by the same amount.}
\label{fig:ast-diam}
\end{figure*}

For deflections less than $\unit[100]{R_{\earth}}$, miss distance scales roughly linearly with $a_\text{l}$ and so scales inversely with mass and therefore the cube of diameter, with the assumed uniform density, as shown in Figure \ref{fig:ast-diam}. A large \unit[100]{N} stand-on mission can deflect a \unit[500]{m} asteroid in an Apophis-like orbit by $\unit[2]{R_{\earth}}$ in under \unit[5]{yr}. Alternatively, a \unit[350]{m} asteroid can be deflected in \unit[3]{yr} or a \unit[170]{m} asteroid in less than \unit[1]{yr}.

When applied over multiple orbits, $0\degree$ thrust tends to delay the asteroid's arrival to the impact point by expanding the orbit of the target, yielding a delay in phase along its orbit. This phase delay opposes the competing effect from the $0\degree$ thrust, which, being a push in the forward direction, tends to speed up the target locally, advancing the asteroid's arrival at Earth when the thrust is applied immediately prior to the encounter. Only when the thrust acts on the asteroid for only a fraction of its orbit does the local speeding effect becomes significant relative to the phase delay, potentially allowing one effect to neutralize the other. If thrust is only possible in the $0\degree$ direction, it should therefore be deactivated before this transition occurs to maximize deflection.

Ideally, direction of thrust should be altered from being parallel to being perpendicular to asteroid's velocity for its final approach to its encounter with Earth. Doing so averts the harmful final speed-up and, instead, has the thrust work to shift the orbit of the asteroid directly. When considering a constant direction of thrust as in these simulations, an ``optimal'' fixed direction may be selected as a weighted average of the ideal thrust direction in each regime.

\begin{figure*}
\plottwo{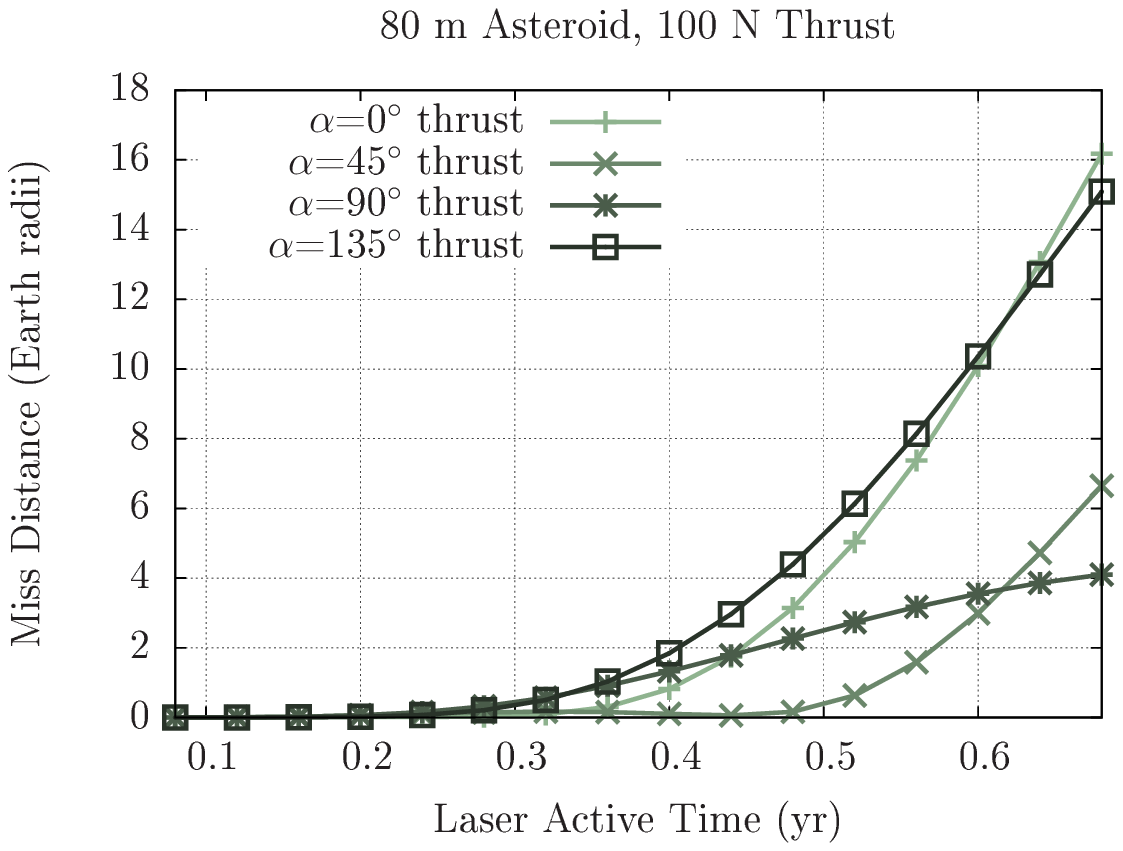}{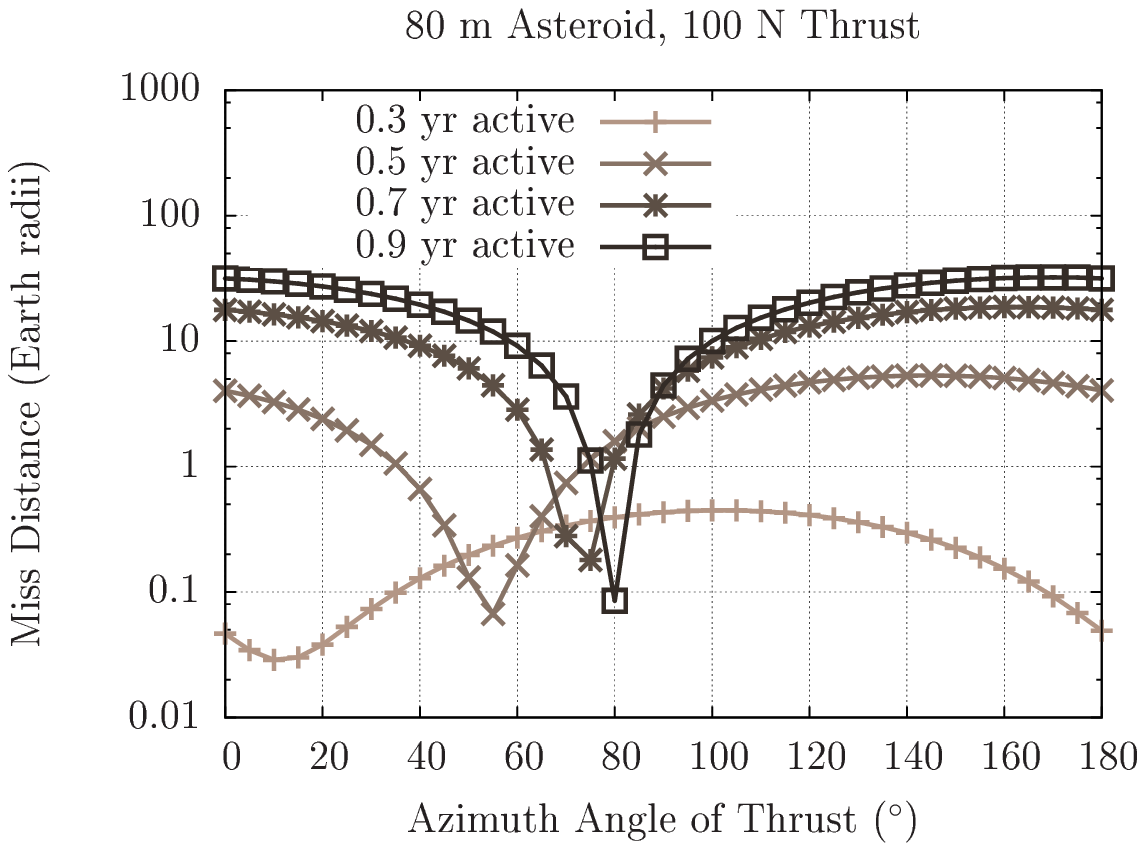}
\plottwo{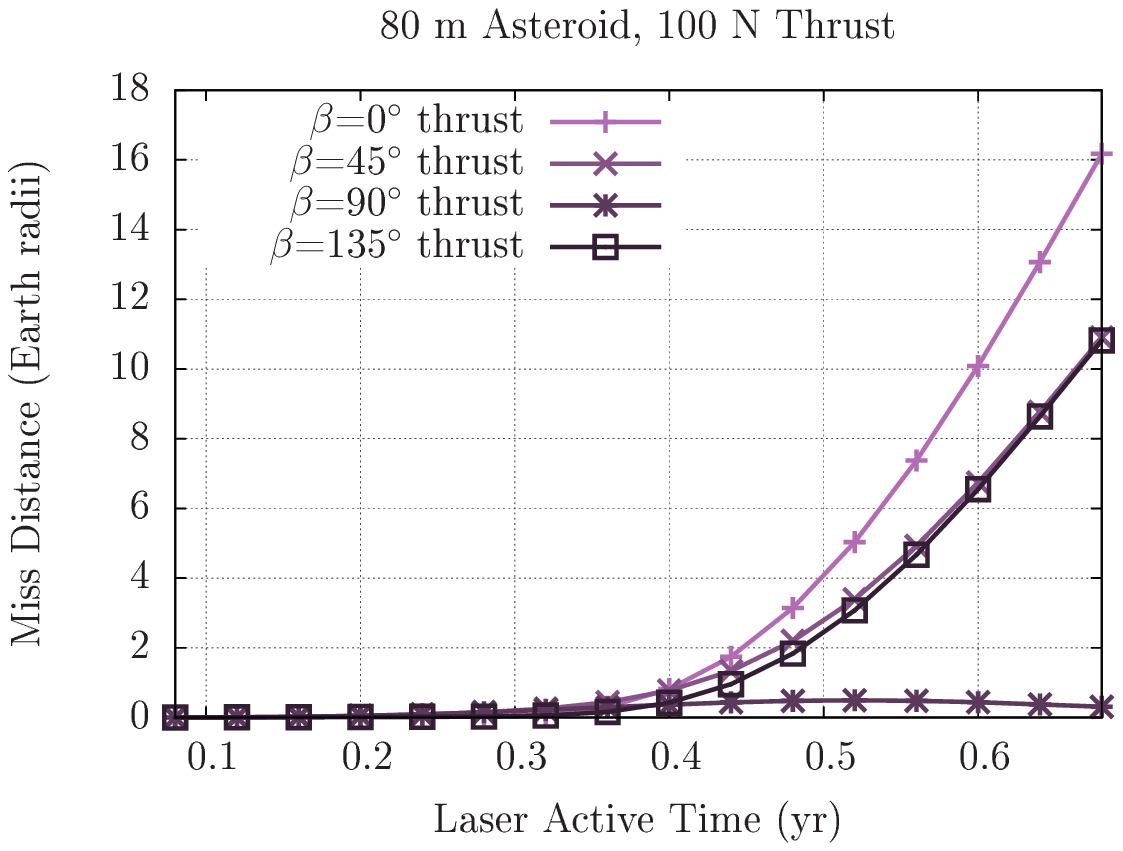}{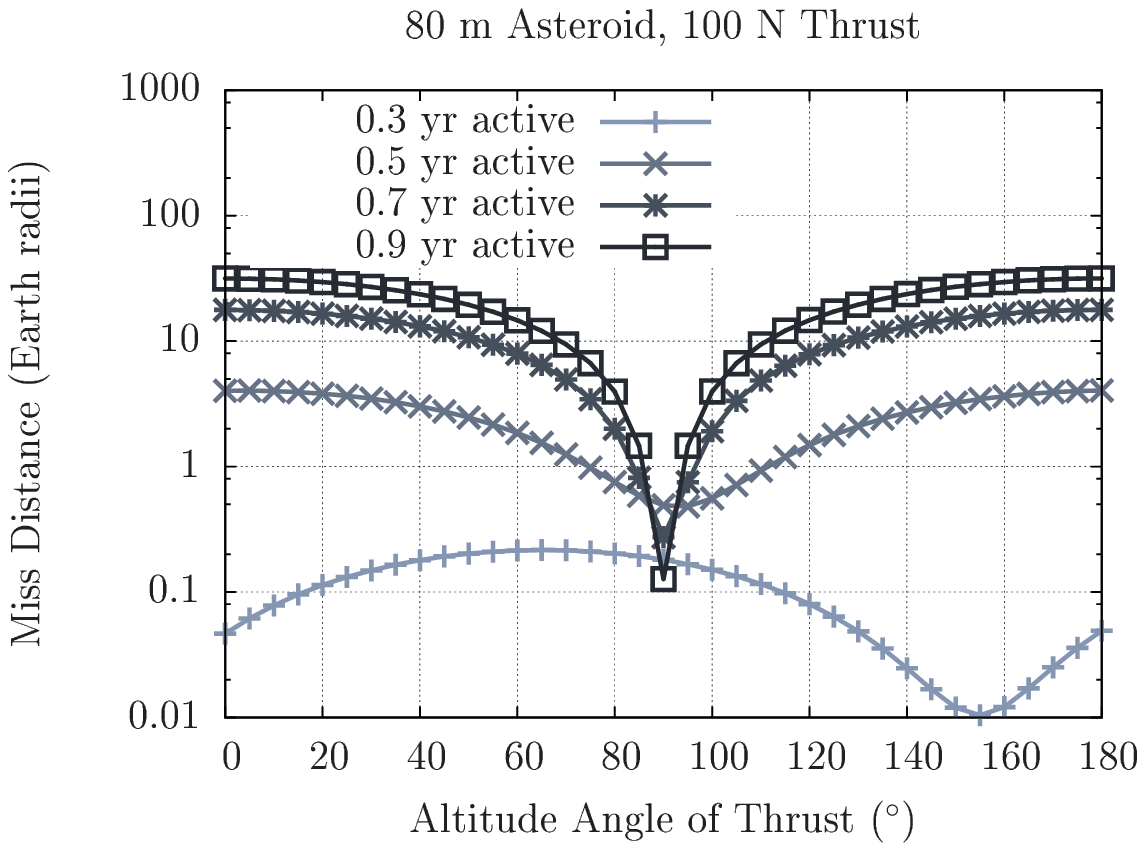}
\caption{A \unit[1]{MW} laser generating \unit[100]{N} thrust acts on an \unit[80]{m} asteroid beginning less than one orbit before the asteroid's encounter with the Earth. Decreasing the duration of laser activity to less than one orbital period causes the optimal direction for thrust to deviate from being parallel to the asteroid's velocity and towards being orthogonal to the velocity. The dependence of deflection distance on duration of laser activity (upper and lower left) is different for different directions of thrust. The shift in optimal thrust direction is already pronounced for a duration of one period ($\unit[\sim0.9]{yr}$) in the azimuth ($\alpha$) direction which is optimal near $\alpha\sim10\degree$ (upper right). A similar shift in the altitude ($\beta$) direction (lower right) is only attained for a laser active over half of a period ($\unit[\sim0.45]{yr}$).}
\label{fig:thrust-dir}
\end{figure*}

As an example, a small \unit[80]{m} asteroid -- roughly the size of the 1908 Tunguska impactor -- can be deflected $\unit[2]{R_{\earth}}$ with \unit[100]{N} thrust in less than \unit[6]{months}. To do so, however, would require a shift away from $0\degree$ thrust. The simulations show, in Figure \ref{fig:thrust-dir}, that as the time on target of the laser decreases, the optimal values of angles $\alpha$ and $\beta$ both shift from $0\degree$ towards $90\degree$. However, while the optimal $\alpha$ begins to shift as laser active time approaches \unit[0.9]{yr} (approximately one orbit of the asteroid), the optimal $\beta$ remains nearly fixed at $0\degree$ until the duration of laser activity drops below \unit[6]{months}. This result is consistent with the notion that shifting the path of the asteroid within the plane of its orbit (with $\alpha=90\degree$) requires significantly less total impulse than shifting the orbital plane itself (with $\beta=90\degree$). Note that the issue of a shifting thrust can be avoided entirely with by having the laser arrive at the asteroid earlier than $\unit[\sim1]{yr}$ prior to its Earth encounter, a comparatively short period considering the expected transit time of several years.

\begin{figure*}
\plottwo{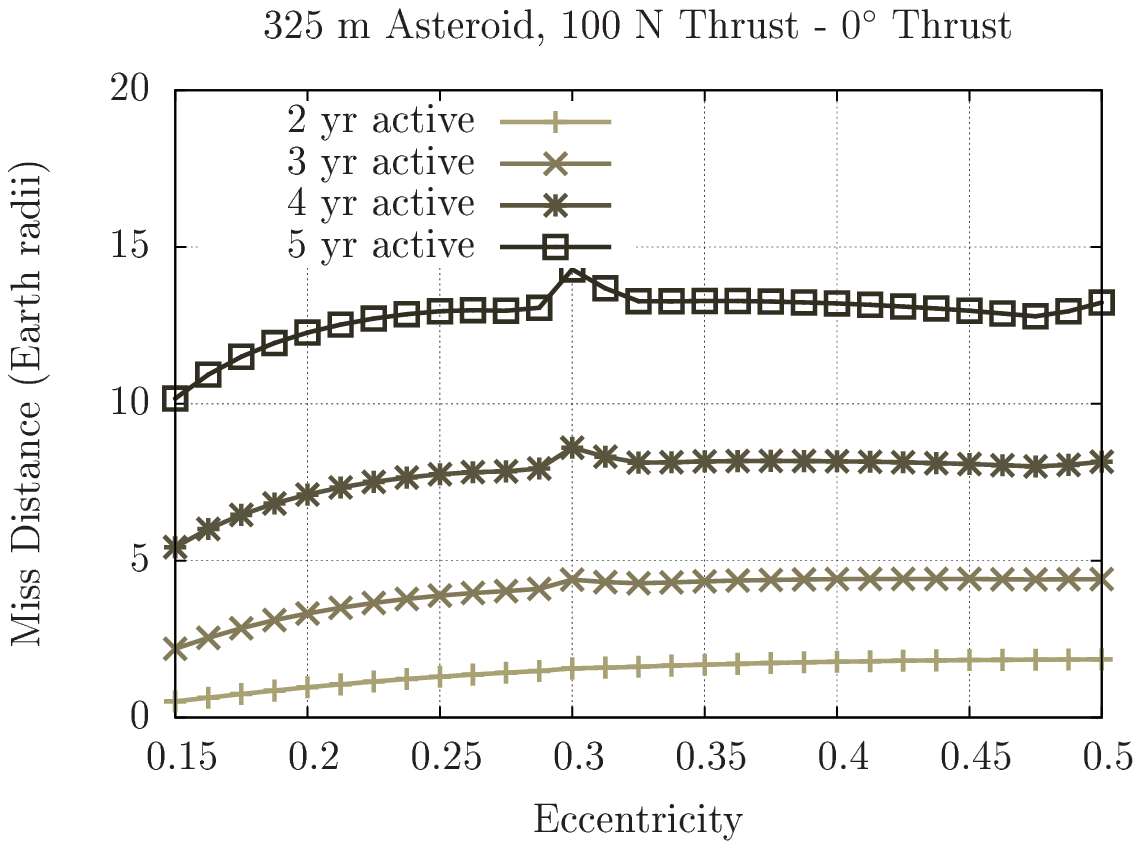}{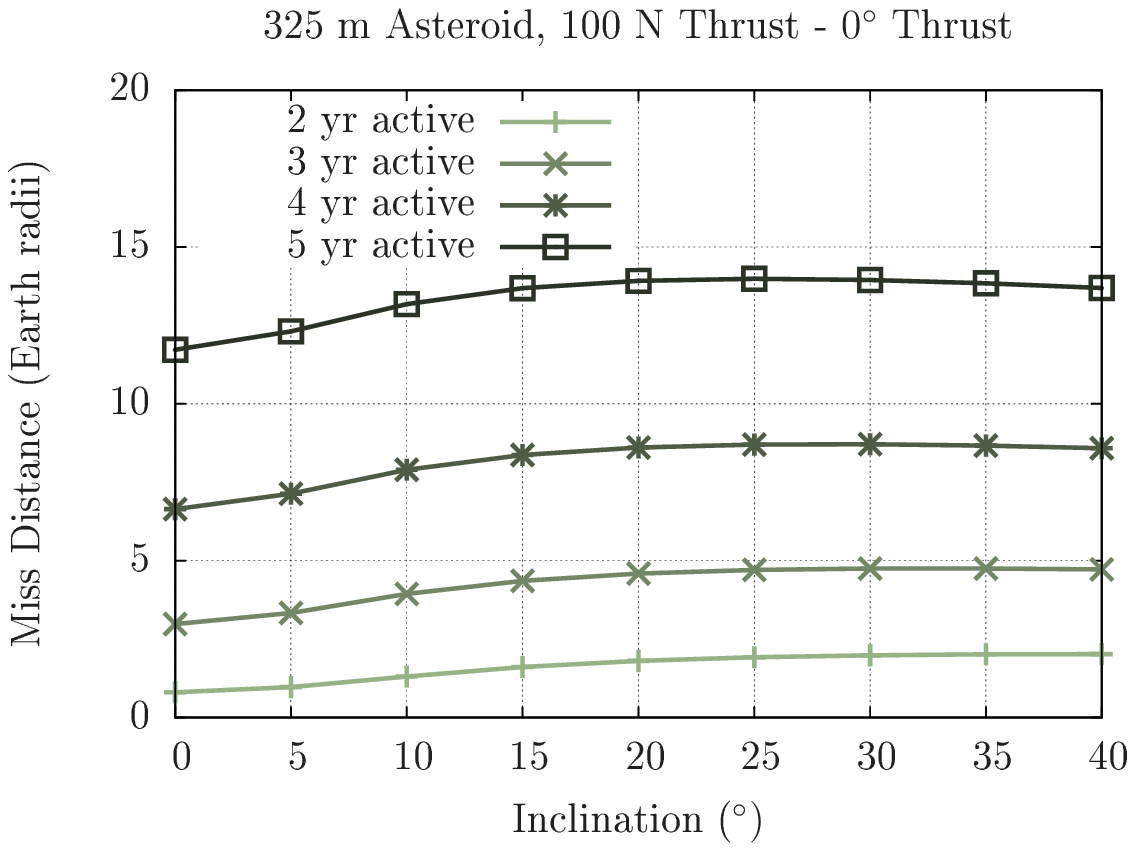}
\caption{Revisiting the case of \unit[100]{N} applied to a \unit[325]{m} asteroid: the orbit of Apophis was taken with its eccentricity and inclination independently varied, and the resulting deflection distances were compared. The miss distance of the asteroid is generally larger for an orbit with higher eccentricity (left) and greater inclination (right) up to a point beyond which deflection distance flattens out. A slight bump in deflection is observed for these cases at $e=0.3$. This bump is the result of a near miss of the asteroid to Earth at $t=T-\unit[1.3]{yr}$ which occurs for this particular set of orbits only when $e=0.3$, an example of a weak keyhole effect where a prior close approach amplifies deflection distance.}
\label{fig:ast-orbit}
\end{figure*}

In addition, the orbit of the target asteroid also affects the effectiveness of thrust in deflecting the asteroid. To measure these effects, a \unit[325]{m} asteroid in an Apophis-like orbit is taken with $e$ and $i$ varied independently in ranges typical of known near-Earth asteroids. Simulations of asteroids in these orbits suggest that for a given amount of thrust, deflection distance grows as the orbit of the asteroid becomes more different from the Earth's -- that in general, larger $e$ and larger $i$ correspond to increased deflection up to a point, beyond which there is little change as seen in Figure \ref{fig:ast-orbit}.

Note, however, that this result does not imply that asteroids with orbits very different from the Earth are easier to deflect with a stand-on mission. Orbits dissimilar to that of Earth's with large $e$ and $i$ require a large $\Delta v$ to reach from Earth. Therefore, the mass and, consequently, the power of the laser that may be delivered to the asteroid will be significantly lower, possibly by several factors, for a given launch configuration \citep{elvis:2011}. Assuming a linear relationship between craft mass and power, the mild gains in deflection per thrust from a highly eccentric and inclined orbit are largely offset by the far much more significant reduction in thrust.

\subsection{Stand-Off Results}

In contrast to a stand-on mission where any laser activity is preceded by a potentially lengthy transit period, a stand-off system may be used as soon as an asteroid is identified as a threat, provided the system is already in place. Being limited by $\Delta_\text{crit}$, stand-on systems are generally restricted to operation over very short timescales on the order of a few days or weeks unless the phased array is at least several kilometers in diameter. Systems of such scale are necessary to deflect larger asteroids like Apophis.

Simulations were run for asteroids in the canonical Apophis-like orbit and indicate that the smallest useful stand-off array to defend against small asteroids ($\unit[\sim20]{m}$) is about $d=\unit[600]{m}$ while a larger $d=\unit[1]{km}$ array may be somewhat effective against Tunguska-class impactors of $\unit[\sim80]{m}$ diameter. Operating at its standard $P=\unit[0.7]{GW}$ (corresponding to $F_0=\unit[70]{kN}$, at $\unit[100]{\mu N\cdot W^{-1}}$), a $\unit[1]{km}$ system can deflect an \unit[80]{m} asteroid vaporizing at $T_\text{crit}=\unit[2\,500]{K}$ by $\unit[0.3]{R_{\earth}}$ over the course of \unit[4]{weeks}. While generally insufficient to prevent an impact, a deflection of this magnitude is more than sufficient to relocate the impact ellipse to a more favorable site.

\begin{figure*}
\plottwo{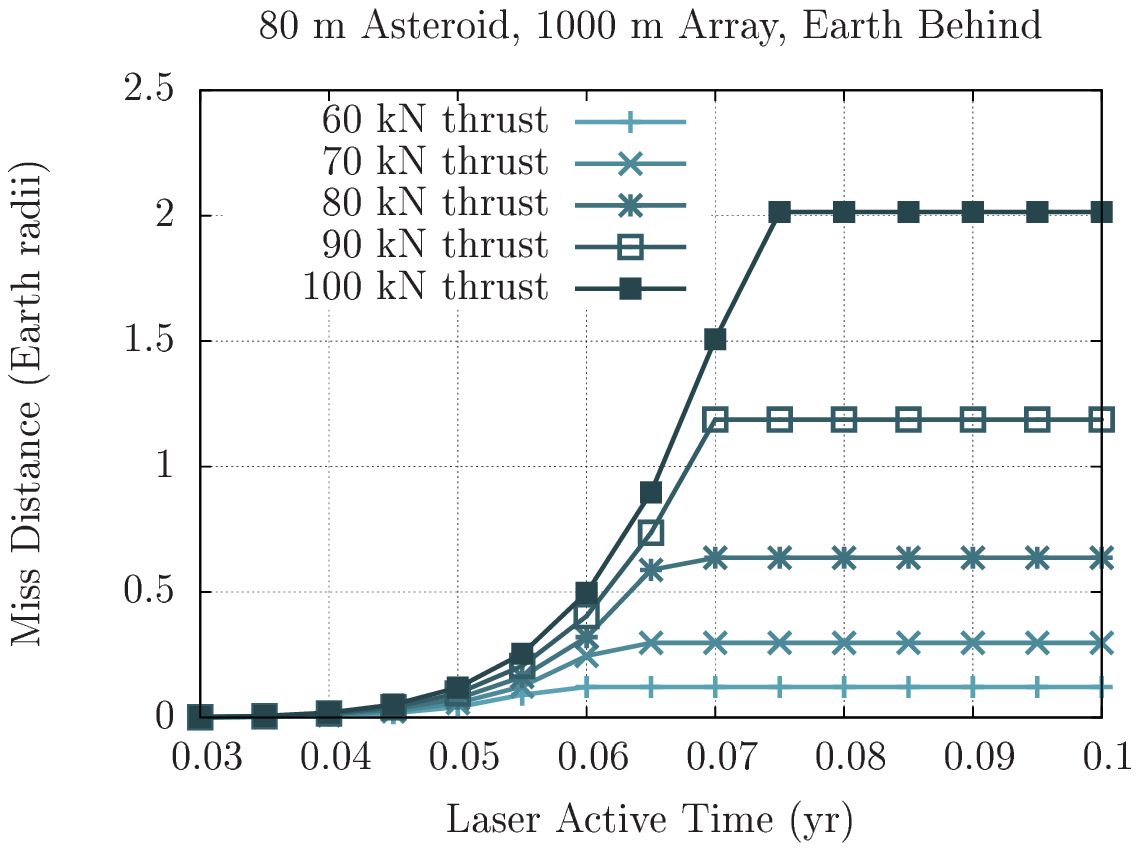}{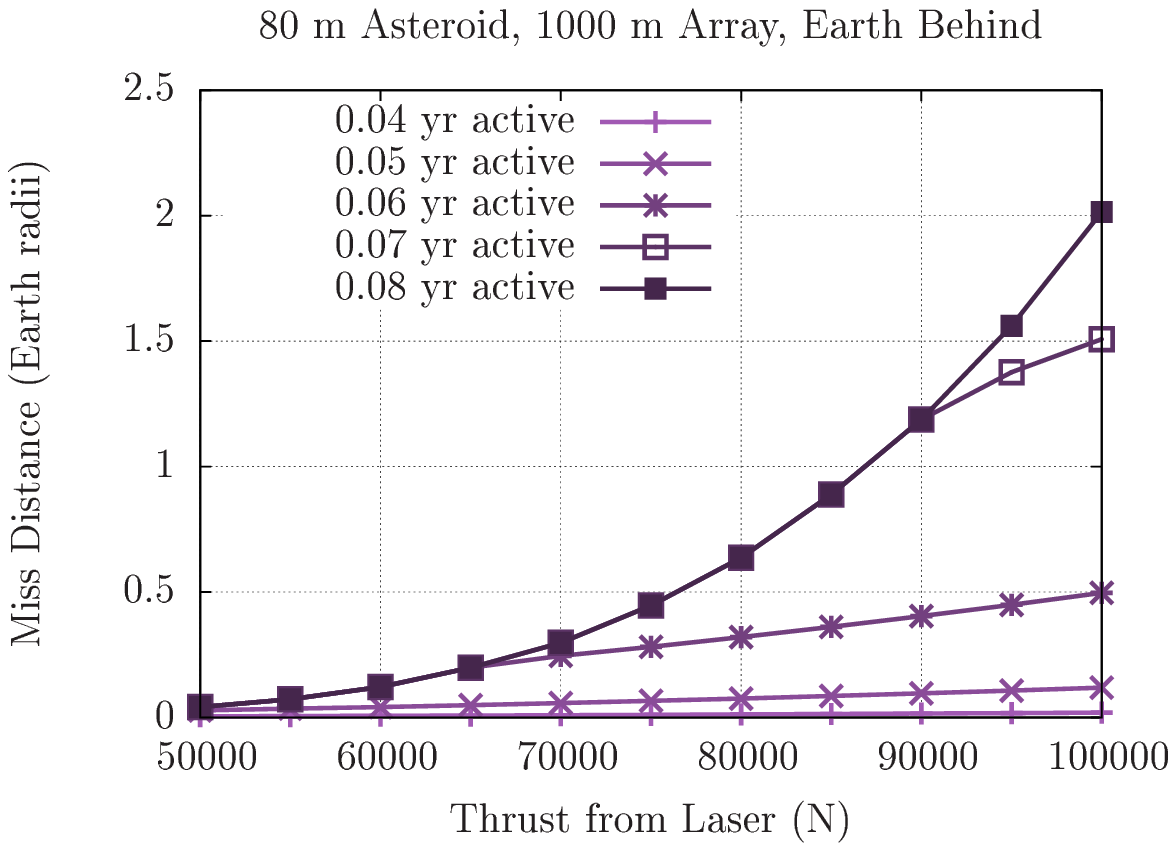}
\caption[]{A \unit[1]{km} phased-array laser in Earth orbit deflects an \unit[80]{m} Tunguska-class asteroid. Deflection distance grows like the stand-on case with laser active time to a certain point before flattening out; activating the laser before the asteroid approaches to within $\Delta_\text{crit}$ yields no additional effect on deflection distance (left). In addition, deflection distance grows roughly quadratically with increased thrust if ablation is begun before the asteroid reaches $\Delta_\text{crit}$ (right). Increased power increases $\Delta_\text{crit}$ and the time over which ablation occurs, and deflection is proportional to the square of the duration of laser activity. Otherwise, linear growth with thrust is observed for cases when the flux is sufficient for ablation to occur over the full period as was observed in stand-on mode. At large thrust/power, $\Delta_\text{crit}$ is larger and so the period over which ablation occurs is longer, eventually covering the entire duration of laser activity. As a result, there is a transition from greater-than-linear growth, a characteristic of varying laser time, to linear growth, a characteristic of constant time, in deflection with thrust.

When powered entirely by a \unit[1]{km} solar array, an efficiency of 50\% corresponds to a power of \unit[700]{MW}, a maximum thrust of \unit[70]{kN} and a deflection of $\unit[0.3]{R_{\earth}}$ which is generally insufficient to completely avert an impact. Such a deflection, however, may be sufficient to relocate the site of impact away from a populated area given a sufficiently well-determined orbit. Increasing the power to \unit[1]{GW}, perhaps with a supplementary battery system, is necessary for a safe deflection of $\unit[2]{R_{\earth}}$.}
\label{fig:standoff-ast}
\end{figure*}

Figure \ref{fig:standoff-ast} shows that deflection distance is strongly dependent on laser power and thrust. A drop in laser power to $P=\unit[500]{MW}$ ($F_0=\unit[50]{kN}$) results in nearly zero deflection. Conversely, increasing laser power to $P=\unit[1]{GW}$ ($F_0=\unit[100]{kN}$), significantly increases the deflection distance to $\unit[2]{R_{\earth}}$ which is sufficient to prevent an impact completely given a well-determined orbit a month in advance. Such high power, while unrealistic for a purely solar-powered laser of this scale, might be possible with the support of a pre-charged battery system or an alternative supplemental energy source. Note that activating the laser before $T-\unit[1]{month}$ (laser active for $\gtrsim\unit[1]{month}$) yields no additional deflection due to the asteroid being out of range ($\Delta>\Delta_{\text{crit}}$) during this time.

\begin{figure}
\plotone{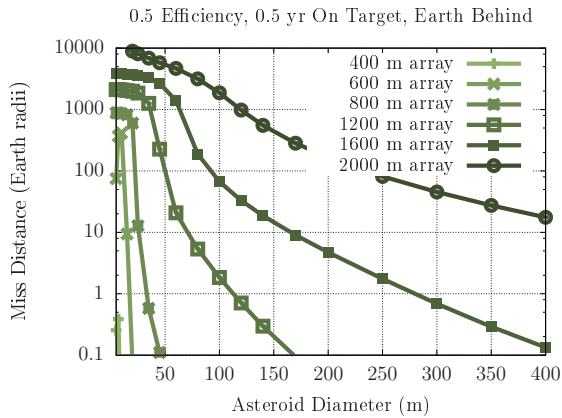}
\caption{Increasing laser/solar array size beyond \unit[800]{m} yields a rapid increase in the effectiveness. In addition to increasing power/thrust, an increase in array size also increases $\Delta_{\text{crit}}$ which permits ablation to begin earlier and occur longer. Operating at 50\% solar-to-laser efficiency, \unit[1.2]{km} array can deflect a \unit[100]{m} asteroid by $\unit[2]{R_{\earth}}$. A \unit[1.6]{km} array, at $1.8\times$ the size of a \unit[1.2]{km} array, can deflect a \unit[250]{m} asteroid -- $16\times$ the mass of a \unit[100]{m} asteroid -- by the same distance. An even larger \unit[2]{km} array can mitigate all probable near-term threats, being capable of deflecting even a large \unit[400]{m} asteroid by a very comfortable $\unit[20]{R_{\earth}}$. Conversely, a \unit[400]{m} array is insufficient to deflect an asteroid of any size under purely solar power.}
\label{fig:ast-standoff}
\end{figure}

Increasing array size beyond \unit[1]{km} rapidly increases the size of asteroid that can be deflected. Increasing array size increases power $P$ and decreases the laser beam divergence angle and thus spot size $D_{\text{s}}$. Both effects contribute to an increase in $\Delta_{\text{crit}}$, extending the duration of time for which the laser may be active, which, when coupled with the increased $F_0$, produces an extremely strong dependence of deflection on array size.  Figure \ref{fig:ast-standoff} shows the effectiveness of various arrays operating on solar power at 50\% efficiency. With a \unit[2]{km} array, even very large asteroids of \unit[400]{m} diameter can be deflected by a more-than-sufficient $\unit[20]{R_{\earth}}$.

\subsubsection{Comet Deflection}

Long-period comets pose a risk frequently neglected in most studies of impact avoidance schemes including most of those listed in the introduction of this paper. This section does not intend to provide a comprehensive analysis of directed energy comet deflection. A proper treatment of the comet deflection problem demands a model substantially more complicated than the linear $\unit[100]{\mu N\cdot W^{-1}}$ model presented here for asteroids. Such treatment must consider the substantial variations in heating response from comet-to-comet, even those of a similar class, often by factors of 10 or more \citep{yeomans:2004}. Rather, this section intends only to take a cursory look into how comet deflection \emph{might} be done under a select few scenarios that could be encountered in reality. For this intent, the model developed for the asteroid simulations suffices.

Im general comets form a difficult class of targets to target due in part to the nature of their orbits. A typical long-period comet with $e=1$ would approach the Earth at a relative speed of $v_{\text{rel}}\equiv\|\vec{v}-\vec{v}_{\earth}\|$ constrained by

\begin{equation}
\begin{aligned}
(\sqrt{2}-1)\sqrt{\frac{GM_{\sun}}{\unit[1]{au}}} &< v_{\text{rel}} < (\sqrt{2}+1)\sqrt{\frac{GM_{\sun}}{\unit[1]{au}}}\\
\implies\unit[12.4]{km\cdot s^{-1}}&<v_{\text{rel}}<\unit[71.9]{km\cdot s^{-1}}
\end{aligned}
\end{equation}

To directly intercept the comet and match its orbit, a $\Delta v\gtrsim v_{\text{rel}}$ is necessary given the typically short time frame $\lesssim\unit[2]{yr}$ between discovery and perihelion and thus, to good approximation, perigee \citep{francis:2005}. In the absence of a propulsion mechanism capable of the high $\Delta v$ needed for typical Earth-crossing long-period comets, such targets are inaccessible to stand-on missions. However, unlike a stand-on system which must be physically delivered to its target, a stand-off system in Earth orbit can target and provide thrust to objects approaching the Earth in any direction, including fast-moving comets.

Unlike asteroids, comets are already being heated by solar radiation to a temperature where its ices are already vaporizing. This behavior increases the difficulty of predicting a comet's trajectory and thus determining whether the comet is a threat. An additional consequence is that the range of ablation extends to the entire zone around the Sun in which a comet will display cometary behavior, a condition that must usually be satisfied for the comet to be sufficiently bright for discovery. The energy from the laser beam supplements the received solar energy and contributes an additional perturbation to its trajectory, potentially deflecting it from an otherwise collisional trajectory.

\begin{figure}
\plotone{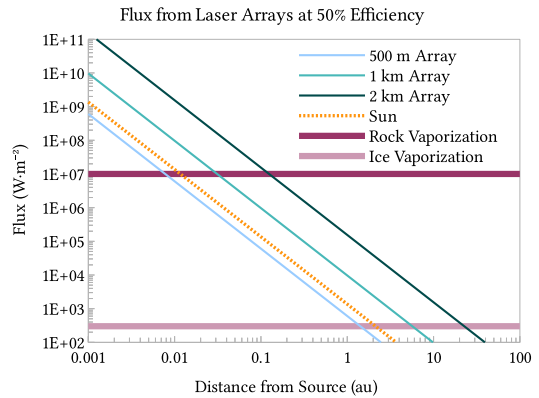}
\caption[]{Rocky material on most asteroids vaporizes when the flux is above $\unit[\sim10]{MW\cdot m^{-2}}$ (dark horizontal bar) while water ice, as found on comets, vaporizes at a much lower $\unit[300]{W\cdot m^{-2}}$ (light horizontal bar). Laser flux (diagonal solid blue lines) falls off with the square of distance from the laser, giving a distance limit -- $\Delta_\text{crit}$ -- beyond which the flux is too low to vaporize material. Due to the significantly lower flux needed to vaporize water ice compared to rock, the $\Delta_\text{crit}$ for ablating ice off the surface of a comet is 200 times larger than the $\Delta_\text{crit}$ for vaporizing rock on an asteroid. With comets, even the Sun (diagonal dotted orange line) has a profound effect as its $\Delta_\text{crit}$ for water ice extends to \unit[2]{au}. Any comet passing within that distance from the Sun already receives sufficient flux from the Sun to vaporize water ice. Additional flux from a laser will add to the thrust already generated by the Sun producing a deflection from the comet's natural trajectory. A large \unit[2]{km} array can extend the zone where water ice vaporizes to over \unit[20]{au}, the orbit of Uranus.}
\label{fig:flux}
\end{figure}

Figure \ref{fig:flux} illustrates how flux declines with distance for \unit[500]{m}, \unit[1]{km} and \unit[2]{km} arrays compared with the flux from the Sun. The flux needed to vaporize typical basaltic rocks and water ice are also included to show the maximum range -- $\Delta_\text{crit}$ -- at which each source can ablate the surface an asteroid and water ice on a comet. The $\Delta_\text{crit}$ for water ice from the Sun alone extends to \unit[2]{au}. Any comet passing within that distance of the Sun -- which is necessary for the comet to be a threat to Earth -- already receives sufficient flux from the Sun to vaporize water ice, hence the cometary behavior. In addition, many comets -- especially dynamically new comets -- have a surface covered by significant fraction of other volatiles that vaporize at even lower fluxes and so have even larger $\Delta_\text{crit}$. For these simplified orbital simulations, the comet is assumed to have been discovered while active and thus receives sufficient flux from the Sun alone for vaporization to occur.

Simulations were run for various-sized comets with an orbit with perihelion $q=\unit[0.8]{au}$, eccentricity $e=0.98$ and inclination $i=130\degree$. Thrust from the laser is assumed to be radial from the Earth as with the asteroid cases, similarly falling off with $1/\Delta^2$ when the beam size exceeds the size of the comet.

Note that although the effect of heating from the Sun may be more significant than the effect on heating by the laser, solar heating only contributes to the natural trajectory of the comet. The goal here is not to analyze the perturbations from a purely gravitational trajectory, but rather, the perturbations from the natural trajectory. Assuming a linear relationship between power and thrust, the perturbations by the laser and by the Sun will obey the law of superposition (for perturbations much smaller than the force of gravity), permitting the two effects to be considered independently.

\begin{figure}
\plotone{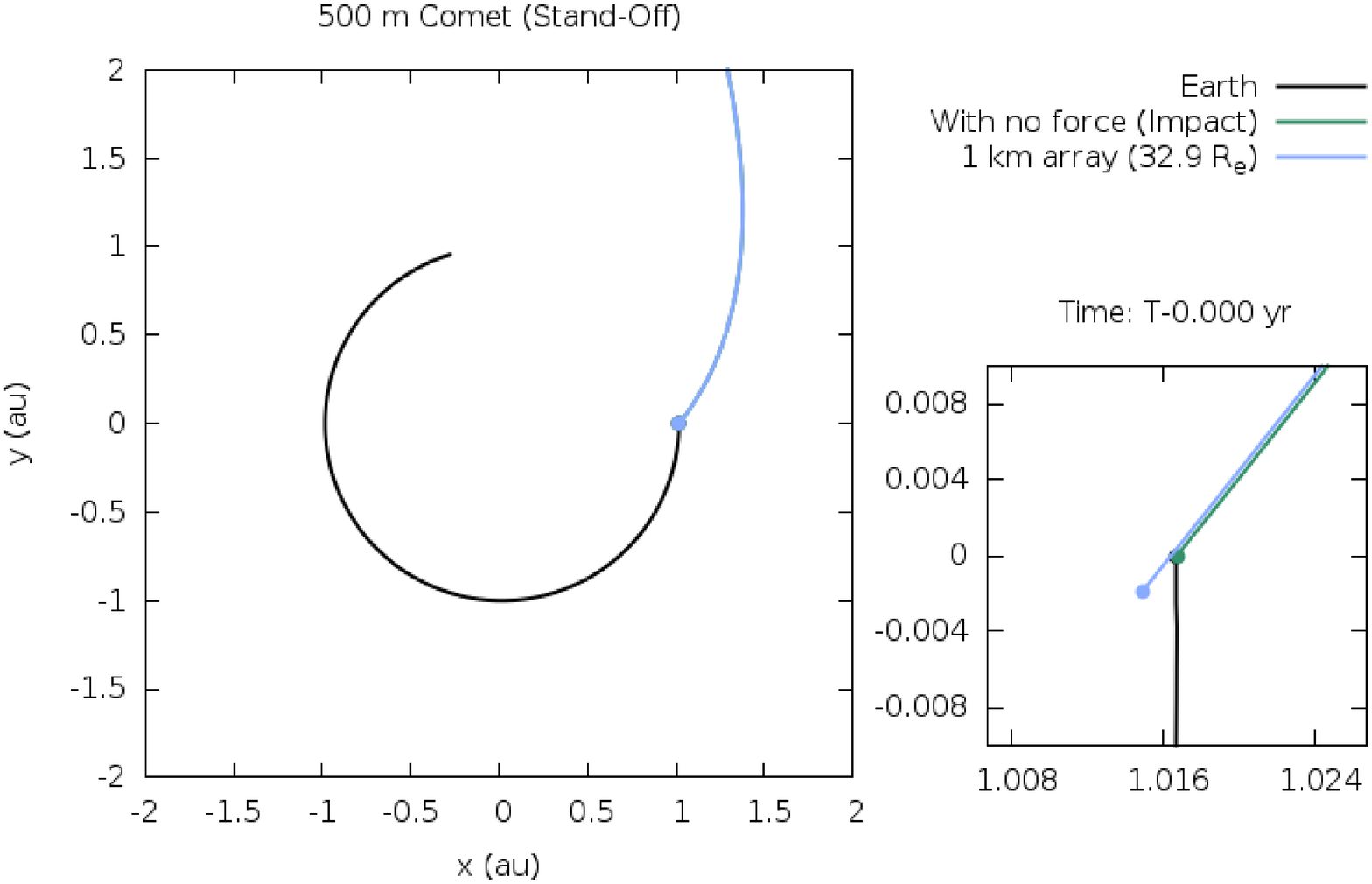}
\caption{Long period comets passing through the inner solar system often pass perihelion (and thus the Earth) in under \unit[2]{yr} after discovery. Such short notice coupled with the highly eccentric and inclined orbits of many of these comets makes the delivery of a stand-on system to a threatening comet infeasible. For such targets, a stand-off system is the only possibility for deflection with directed energy. In this figure, a \unit[500]{m} comet with $e=0.98$ and $i=130\degree$ (green) approaches from above and impacts the Earth (black) approaching from below in a near head-on collision. Activating a \unit[1]{km} stand-off array \unit[2]{yr} in advance leads the deflected comet (light blue) to arrive at the Earth's orbit before the Earth, averting the impact as evident in the inset (lower right).}
\label{fig:500m-com}
\end{figure}

\begin{figure}
\plotone{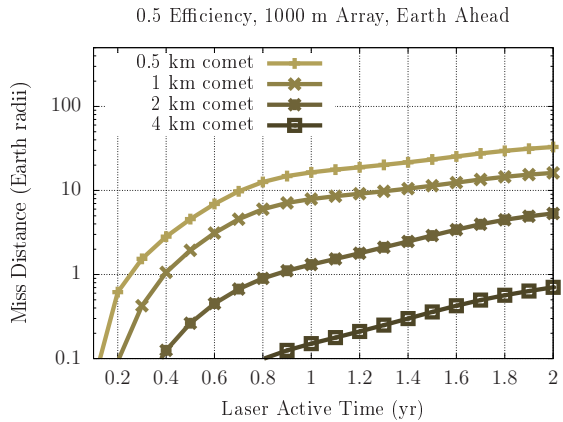}
\caption{A \unit[1]{km} laser array at 50\% efficiency produces \unit[70]{kN} thrust which can deflect a comet as large as \unit[2]{km} by $\unit[5]{R_{\earth}}$ with \unit[2]{yr} of warning. A more common \unit[500]{m} comet can be deflected by a much larger $\unit[30]{R_{\earth}}$, a distance generally sufficient to overcome the anticipated uncertainties in the comet's computed trajectory.}
\label{fig:com-deflect}
\end{figure}

For these simulations, a constant conversion factor of $\unit[100]{\mu N\cdot W^{-1}}$ is used, the same factor as the one used for asteroids. Figure \ref{fig:500m-com} illustrates the orbital deflection of a \unit[500]{m} comet by a \unit[1]{km} array. Figure \ref{fig:com-deflect} shows that with the \unit[1]{km} array, a \unit[500]{m} comet may be deflected by $\unit[30]{R_{\earth}}$ or a \unit[2]{km} comet by $\unit[5]{R_{\earth}}$ given \unit[2]{yr} of warning.

\begin{figure*}
\plottwo{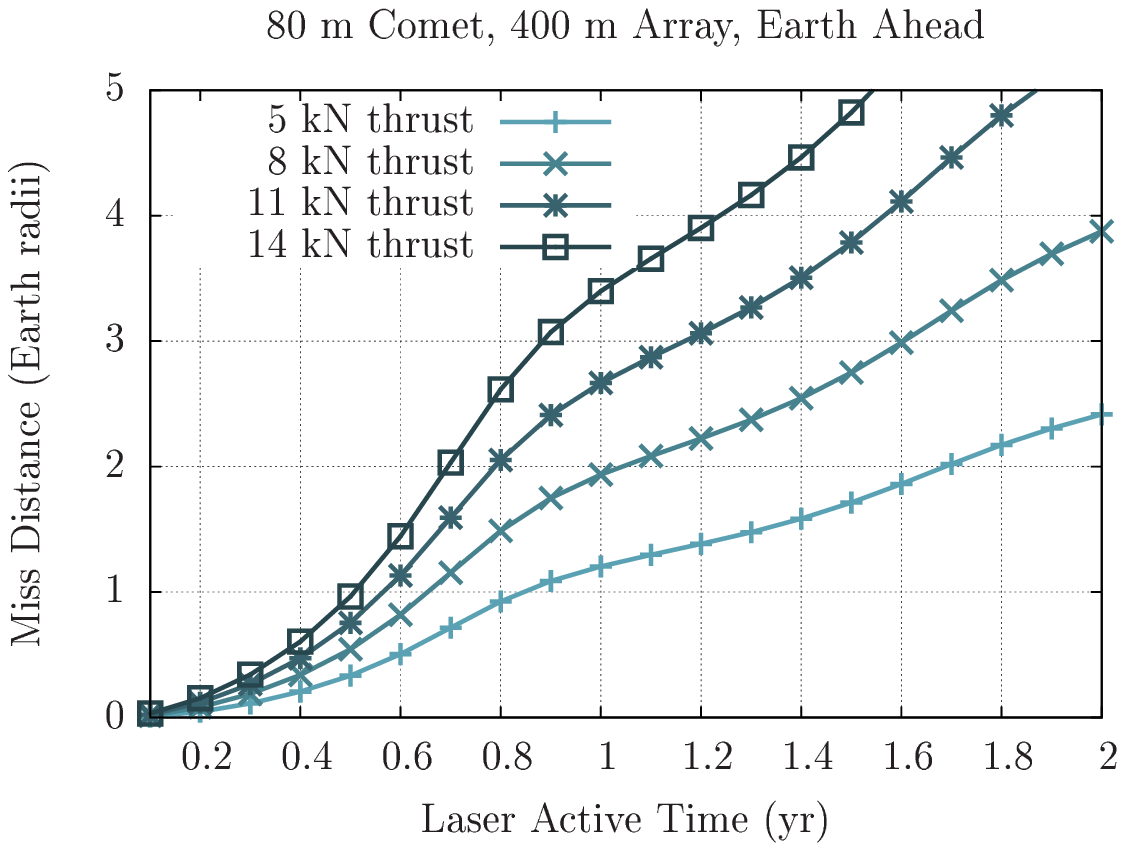}{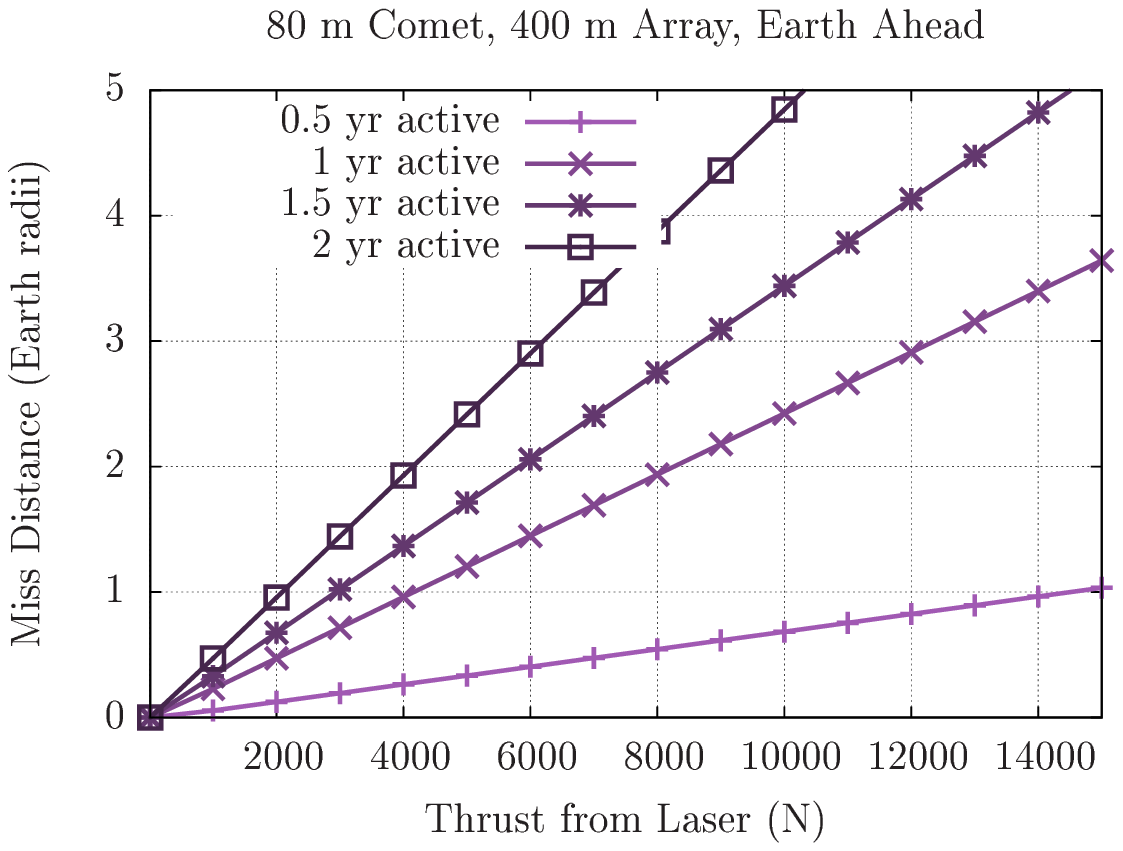}
\caption{Deflection of an 80 m comet by a \unit[400]{m} stand-off laser array: At 50\% solar-to-laser efficiency, the array produces a combined beam of \unit[110]{MW} optical power yielding a maximum thrust of \unit[11]{kN} at $\unit[100]{\mu N\cdot W^{-1}}$. Miss distance increases quadratically with increasing laser active time to a certain point. Starting the laser even earlier results in a period where the laser spot diameter $D_\text{spot}$ is bigger than the diameter of the comet's nucleus $D$ which still contributes towards deflection (unlike the asteroid case which flattens out after the spot size exceeds $D_\text{crit}$) but at a rate slower than quadratically (left). For a fixed laser array size, increasing efficiency and therefore thrust increases the miss distance of the comet linearly (right).}
\label{fig:80m-com}
\end{figure*}

The effectiveness drops rapidly as the arrays are scaled down. The minimum size of an array of use in deflecting comets is about \unit[400]{m}. Such an array in Earth orbit operating at 50\% efficiency yields $F=\unit[11]{kN}$. This \unit[400]{m} / \unit[11]{kN} system can deflect a small \unit[80]{m} comet by $\unit[2]{R_{\earth}}$ in one year, or $\unit[5]{R_{\earth}}$ in two. As shown in Figure \ref{fig:80m-com}, increasing the laser active time $T$ increases deflection distance roughly quadratically for small $T$. For larger $T$, laser activity begins before the comet is sufficiently close to intercept the entire laser beam resulting in a deflection distance that increases slower than quadratically. Improving laser efficiency and thus increasing the thrust exerted on the comet scales with deflection distance linearly just as with the stand-on cases.

The simulations show that stand-off arrays smaller than $\unit[\sim350]{m}$ are unlikely to be able to deflect a comet or asteroid of any size. These smaller arrays may still be useful in mitigating very small threats ($\unit[\sim20]{m}$) by vaporization or otherwise total disintegration. Structural  analysis, however, is beyond the scope of these simulations which assume a target of constant mass.

Generally, a stand-off system is significantly less effective than a similarly-sized stand-on mission. Due to the divergence of the laser beam over large distances by diffraction effects, a very large laser array of at least \unit[1]{km} is needed to concentrate enough flux into a spot to ablate material off an asteroid sufficiently far away for a significant deflection. The lack of a transit time, however, makes a stand-off setup the only directed energy option for deflecting incoming asteroids on short notice. Furthermore, such stand-off systems are significantly more effective on long-period comets. These targets may approach in a trajectory unreachable by modern spacecraft propulsion systems making such a system stand out as one of the very few options available to mitigate such threats. With either case, a stand-off system needs a large array of at least several hundred meters to be effective at deflecting any target and thus remains a long-term option rather than an immediate solution.

\section{Conclusions}

Directed energy is a promising technology for planetary defense. A modest stand-on DE-STARLITE mission of just \unit[1]{MW} (\unit[100]{N}), which fits within a single SLS Block 1 launch configuration, can deflect all known threats up to \unit[500]{m} in diameter with \unit[5]{yr} of laser activity. That same system could deflect Tunguska- or Chelyabinsk-sized asteroids in well under a year upon arrival at the asteroid. With the strong dependence of deflection on laser active time, a much smaller and less expensive system could be equally effective given a decade or more of activity. Conversely, stand-on systems are largely ineffective at deflecting targets on short notice due to the time required for transit to the target asteroid.

In the absence of more than a few weeks of warning, a very large stand-off DE-STAR system becomes the only option. In addition to providing a last line of defense against threats which have evaded detection until immediately before impact, such a system may also provide one of the few options for defense against long period comets to which modern technology is often incapable of reaching by spacecraft. With the support of a battery system, the ablation range and thus effectiveness for a stand-off system might conceivably be extended by a few factors. Even so, a system of sufficient scale will likely require decades to construct and so becomes a possibility only in the more distant future.

The actual effectiveness of a deflection mission depends strongly on the target to be deflected. A mission optimized for one target may be ineffective when applied to another, even one of the same size and composition. Orbital simulations provide a means for determining the specific mission requirements for targeting each specific threat. Planning, however, must begin long before an actual threat is identified. With orbital simulations, classes of threats can be identified and planned for ahead of time, minimizing the build out time thus maximizing the effectiveness of the system upon confirmation of an actual threat.

\acknowledgments
We gratefully acknowledge funding from the NASA California Space Grant NNX10AT93H and from
NASA NIAC 2015 NNX15AL91G. This work incorporates material presented at the 13\textsuperscript{th} Hypervelocity Impact Symposium \citep{zhang:2015:hvis} and at SPIE Optics + Photonics 2015 \citep{zhang:2015:spie-op:4}.

\end{document}